\documentclass[12pt,preprint]{aastex}
\def\chandra{{\it Chandra }}
\begin{document}
\title{Chandra Deep Field South: The 1Msec Catalog}

\author{Riccardo Giacconi\altaffilmark{1, 2}, Andrew
Zirm\altaffilmark{1}, JunXian Wang\altaffilmark{1}, Piero
Rosati\altaffilmark{3}, Mario Nonino\altaffilmark{4}, 
Paolo Tozzi\altaffilmark{4}, Roberto
Gilli\altaffilmark{1, 5}, Vincenzo Mainieri \altaffilmark{3,6}, 
Guenther Hasinger\altaffilmark{7}, Lisa Kewley\altaffilmark{8}, 
Jacqueline Bergeron\altaffilmark{3},
Stefano Borgani\altaffilmark{9}, Roberto Gilmozzi\altaffilmark{3}, 
Norman Grogin \altaffilmark{10},
Anton Koekemoer\altaffilmark{10}, Ethan Schreier\altaffilmark{10},
Wei Zheng\altaffilmark{1} and Colin Norman\altaffilmark{1,10}}

\altaffiltext{1}{The Johns Hopkins University, Homewood Campus,
Baltimore, MD 21218}

\altaffiltext{2}{Associated Universities, Inc. 1400 16th Stret, NW,
Suite 730, Washington, DC 20036}

\altaffiltext{3}{European Southern Observatory,
Karl-Schwarzschild-Strasse 2, Garching, D-85748, Germany}

\altaffiltext{4}{Osservatorio Astronomico, Via G. Tiepolo 11, 34131
Trieste, Italy}

\altaffiltext{5}{Osservatorio Astrofisico di Arcetri, Largo E. Fermi 5,
I-50125 Firenze, Italy}

\altaffiltext{6}{Dipartimento di Fisica ``E.Amaldi'', Universit\'a degli
Studi RomaTre, Via della Vasca Navale 84, I-00146 Roma, Italy}

\altaffiltext{7}{Astrophysikalisches Institute Potsdam, An der
Sternwarte 16, Potsdam, D-14482, Germany}

\altaffiltext{9}{Harvard Smithsonian Center for Astrophysics, 60
Garden Street, Cambridge, MA 02138}

\altaffiltext{9}{INFN, c/o Dip. di Astronomia dell'Universit\`a, via
Tiepolo 11, I--34131, Trieste, italy}

\altaffiltext{10}{Space Telescope Science
Institute, 3700 San Martin Drive, Baltimore, MD 21218}

\begin{abstract}

In this Paper we present the source catalog obtained from a 942 ks
exposure of the \chandra Deep Field South (CDFS), using the Advanced
CCD Imaging Spectrometer (ACIS--I) on the \chandra X--ray
Observatory. Eleven individual pointings made between October 1999 and
December 2000 were combined to generate the final image used for
object detection.  Catalog generation proceeded simultaneously using
two different methods; a method of our own design using a modified
version of the {\tt SExtractor} algorithm, and a wavelet transform
technique developed specifically for \chandra observations. The
detection threshold has been set in order to have less than 10
spurious sources, as assessed by extensive simulations. We subdivided
the catalog into four sections. The primary list consists of objects
common to the two detection methods.  Two secondary lists contain
sources which were detected by: 1) the {\tt SExtractor} algorithm
alone and 2) the wavelet technique alone.  The fourth list consists of
possible diffuse or extended sources. The flux limits at the aimpoint
for the soft (0.5--2 keV) and hard (2--10 keV) bands are
5.5$\times10^{-17}$ erg s$^{-1}$ cm$^{-2}$ and 4.5$\times10^{-16}$ erg
s$^{-1}$ cm$^{-2}$ respectively.  The total number of sources is 346;
out of them, 307 were detected in the 0.5--2 keV band, and 251 in the
2--10 keV band.  

We also present optical identifications for the catalogued sources.
Our primary optical data is $R$ band imaging from VLT/FORS1 to a depth
of $R \sim 26.5$ (Vega).  In regions of the field not covered by the
VLT/FORS1 deep imaging, we use $R$--band data obtained with the Wide
Field Imager (WFI) on the ESO--MPI 2.2m, as part of the ESO Imaging
Survey (EIS), which covers the entire X--ray survey.  We found that
the FORS1/Chandra offsets are small, $\sim 1\arcsec$.  Coordinate
cross-correlation finds $85\%$ of the \chandra sources covered by
FORS1 $R$ to have counterparts within the 3$\sigma$ error box
($\gtrsim1.5\arcsec$ depending on off--axis angle and
signal--to--noise). The unidentified fraction of sources,
approximately $\sim$ 10--15\%, is close to the limit expected from the
observed X--ray flux to $R$--band ratio distribution for the
identified sample.

\end{abstract}

\clearpage

\section{Introduction \label{intro}}

The X--ray background (hereafter XRB) was first identified by
\citet{giacconi1962} and its resolution into discrete sources has
since been a major goal of X--ray astronomy. Deep surveys by the major
X--ray facilities UHURU \citep{matilsky1973}, HEAO-1 A2
\citep{piccinotti1982}, Einstein \citep{giacconi1979}, ROSAT
\citep{hasinger1998}, ASCA \citep{ueda1999} and BeppoSAX
\citep{giommi2000} have resolved an increasing fraction of the XRB in
the 0.5--2 keV and in the 2--7 keV band and produced catalogs of
sources which were then used for a wide variety of other astrophysical
investigations. These surveys were primarily limited by effective area
and source confusion at the faintest achievable fluxes. Unlike these
previous missions, \chandra \citep{weiss2000} provides arcsecond
resolution over a majority of the detector, in particular the Advanced
CCD Imaging Spectrometer \citep[ACIS--I, ][]{garmire1992, bautz1998}.
Superior spatial resolution coupled with a substantial gain in
light--gathering power over ROSAT has allowed our 942 ks exposure
(hereafter 1Msec) of the CDFS (and the analogous exposure of the
Hubble Deep Field North) to become the deepest X--ray exposure(s) ever
taken, improving by factors of $\sim20$ and $\sim200$ the deepest
ROSAT and ASCA surveys, respectively.

Along with the \chandra Deep Field North (CDFN)
\citep{hornschemeier2000,hornschemeier2001, brandt2001a,brandt2001b},
the CDFS provides a unique dataset in which to investigate both
statistical and source-by-source properties of active galaxies over a
large range of redshift and parameter space. Already, the CDFS has
discovered both single interesting objects \citep{norman2001} and
statistical correlations (Giacconi et al. 2001; Tozzi et al. 2001,
hereafter Paper I and II).

The \chandra Deep Field South is centered on $\alpha$ = 03:32:28.0,
$\delta$=-27:48:30 (J2000), and was selected as having: (1) low
Galactic neutral hydrogen column (N$_H$ $\sim$ 8 $\times$ 10$^{19}$
cm$^{-2}$); (2) no bright stars ($m_{\textrm v} \leq 14$) within
30$\arcmin$ and (3) field accessibility from the new 8m class
telescopes, namely VLT and Gemini-South.  Note that recent
higher--resolution HI maps (from the Parkes Multi--beam Survey;
L. Staveley--Smith, private communication) confirm the HI hole in the
CDFS, but do show some sub--structure (on $\sim 10$ arcminute scale)
near the \chandra pointings, at the level of $10^{19.5}$ cm$^{-2}$.

Here we present the catalog of the sources in the CDFS field found in
the 1Msec exposure along with fluxes (or upper limits) in $R$, for
their presumed optical counterparts. The first scientific results
derived from the 1Msec observation are presented in
Rosati et al. 2001 (Paper III).

The Paper is structured as follows: in Section \ref{data} we discuss
the data, in Section \ref{detection} we discuss extraction for point
and extended sources and their photometry; in Section \ref{opticalIDs}
the optical identifications are presented.  We conclude with a
discussion of this dataset.  The common catalog is given in
Table~\ref{common}.  In Tables \ref{sex}, \ref{wav} and \ref{extended} are listed sources detected
only by SExtractor and wavelet techniques, and extended properties
respectively.

\section{X--ray Data \label{data}}

\subsection{Diary of Observations}

The final X--ray image used to generate the catalog is a combination
of 11 individual \chandra ACIS--I pointings summarised in
Table~\ref{ObsTab}. The maximum total exposure time is 942 ks, but
varies across the detector(s), with a minimum of 25 ksec at the edge
of the field where we obtained only single exposures due to rotation
of the field--of--view.  ACIS consists of ten CCDs, distributed in a
2x2 array (ACIS--I) and a 1x6 array (ACIS--S, see the Chandra
Observatory Guide at http://asc.harvard.edu/udocs/docs/docs.html).
All four ACIS--I chips and the ACIS--S3 chip were used for the CDFS
observations.  The telescope aimpoint was centered on the ACIS--I3
chip for each exposure. Effective area and spatial resolution of the
telescope vary inversely as off--axis angle. Unfortunately the S3 CCD
is far enough off--axis to decrease its effective area by a factor of
0.7 (at 4.5 keV) and its spatial resolution by a factor of 0.1 or less
relative to the aimpoint. We have therefore chosen to ignore any data
in the S3 CCD for the purpose of this work.

Our first two observations in 1999 (1431--0, 1431--1, see
Table~\ref{ObsTab}) were taken with the ACIS--I at -110 C.  All other
observations were taken at -120 C.  The CTI--induced QE loss due to
grade redistribution (see Townsley et al. 2001) was mitigated by the
ten degree change, especially in the hard band.  As an example, the
effective area improved by 5\% at 4.5 keV.  The improvement is higher
at larger energies, where, however, the effective area is much
smaller.  The 7 observations taken in December 2000 were processed by
the \chandra X--ray Center using the most recent calibration files
which were introduced in versions R4CU5UPD11 (and later) of the
processing software.  Those observations taken in 1999 and the summer
of 2000 were also re--processed by the \chandra X--ray Center using
the new calibration files.

\subsection{Processing Standards}

CIAO 2.0.1 (see http://asc.harvard.edu/ciao) was used to reduce all of
the data sets. While calculating exposure maps, the program {\it
asp\_apply\_sim} was used to fix the shifts between exposure maps and
X--ray images. The data were filtered to include only the standard
ASCA event grades 0, 2, 3, 4 and 6. All hot pixels and columns were
removed. We also removed the flickering pixels (caused mainly by
cosmic rays afterglows), defined as the pixels with more than two
contiguous events within 3.3 sec.  Time intervals with background
rates larger than 3$\sigma$ over the average global value were
removed. Due to the high background, about 8800 sec were removed from the
level 2 files; most of them were removed from the first exposure,
which contained the only long--lasting high background flare.  The
final exposure time of 942 ks also reflects corrections for CCD
read--out time and time lost due to bad satellite aspect.

There are small shifts (between 20'' and 1'') between the aimpoints of
each exposure.  We calculated these relative shifts by registering the
brightest sources and combining the eleven observations. The combined
data has the same coordinate system as Obs. ID 2406. We extracted
three images from the total 1Msec data: a soft image (0.5--2.0 keV), a
hard image (2.0--7.0 keV) and the total image (0.5--7.0 keV).  The
images were binned 2 $\times$ 2, which gives an image scale of
0.982\arcsec/pixel. The hard and total bands were cut at 7 keV since
above this energy the effective area of \chandra is decreasing, while
the instrumental background is rising, giving a very inefficient
detection of sky and source photons.  Figures~\ref{image_soft} and \ref{image_hard} 
shows the soft and hard images. Exposure maps for
the soft and hard bands were calculated for each observation, and then
combined using the determined shifts and weighting for the individual
exposure times.  The soft exposure map is shown in
Figure~\ref{expmap}.  A single ACIS--I pointing covers a field of
about 0.08 deg$^2$. Due to the different roll angles of the individual
pointings, the final image covers a total of 0.109 deg$^2$, decreasing
rapidly near the flux limit.  The solid angle as a function of the
effective exposure time is shown in Figure~\ref{eff_time}. The sky
coverage (see Figure~\ref{skycov}) is defined as the solid angle
within which a source with a given X--ray flux can be detected at S/N$
> 2.1$.  MARX simulations verify that our model for the sky coverage
is accurate within a few percent.  The computation of the sky coverage
and the simulations are described in \citet{tozzi2001}.

\section{Detection Techniques \label{detection}}

\subsection{Sextractor \label{sextractor}}

To detect sources in the Chandra Deep Field South we ran a modified
version of the SExtractor algorithm \citep{bertin1996} on the 0.5--7
keV image.  This modified detection algorithm is several orders of
magnitude faster than the wavelet algorithm of \citet*{rosati1995} or
WAVDETECT in the CIAO software
\citep{dobrzycki1999,freeman2001}. Detection parameters (e.g.,
threshold, characteristic object size, and ultimately the
signal--to--noise) were chosen to find very faint sources, while
limiting the number of spurious sources to 10, or approximately $\sim
3$\% of the total sample.  The number of fake sources as a function of
the algorithms' parameters was determined via extensive simulations as
described in \citet{tozzi2001}.

SExtractor was first run on the 0.5--7 keV image to produce an initial
list of candidate sources. SExtractor detection parameters were chosen
as a result of simulations and we adopted a detection threshold of
2.4, with a Gaussian filter with $1.5\arcsec$ FWHM and a minimum area
of 5 connected pixels.  SExtractor parameters were chosen to detect
sources to the faintest limits, thus including a sizeable number of
spurious sources.  Object photometry for each source was then
performed separately in the soft and hard bands and its significance
was determined by measuring the signal--to--noise for each source.
The definition of S/N was that adopted by \citet{tozzi2001}.  Source
counts are measured in a circle of radius $R_s$, where $R_s$ is in
units of pixels and is determined by a function $R_s$ = 2.4$\arcsec$
$\times$ FWHM (with a minimum of 5 pixels).  The FWHM was modeled as a
function of the off--axis angle to reproduce the broadening of the
PSF.  The parabolic fit we adopted has the coefficients $(a_0, a_1,
a_2) = (0.6779,-0.0405,0.0535)$.  The radius $R_s$ reproduces the 95\%
encircled--energy radius as included in the WAVDETECT algorithm.  Note
that the fit that we used for photometry is different from the fit of
the average FWHM of the observed sources that we used later to select
extended sources.

The local background was calculated for each source in an annulus with
outer radius of R$_s$ + 12 pixels ($11.8\arcsec$) and inner radius
R$_s$ + 2 pixels ($1.9\arcsec$), after masking out nearby sources.  We
considered only sources with S/N $\geq$ 2.1 in either the soft or hard
band.  This procedure removed 47 sources from the raw catalog obtained
with SExtractor.  Detailed simulations have shown that a S/N$>$2.1
threshold results in less than 10 spurious sources in the whole
sample.  Our simulations have also shown that aperture photometry
leads to an underestimate of the source count rate by approximately
$4$\% (see panel c of Figure 1 of Paper II).  We have corrected this
photometric bias before converting count--rates into energy fluxes.  We
also notice that our aperture photometry is based on apertures larger
than both the hard and soft PSFs, so that we do not suffer any effect
from the energy dependent nature of the PSF.

Incidentally, we remark that the same procedure run separately in both the
soft and hard bands did not add any new detections.  Conversely, the
combined catalog obtained from the detections in the two separate
bands, was missing several {\sl bona fide} sources found in the
0.5--7 keV image.  This is expected since we are far from being
background limited: the soft background is $(2.6\pm 0.4) \times
10^{-7}$ counts/arcsec$^2$/s, while the hard (2--7 keV) background is
$(4.2\pm 0.5) \times 10^{-7}$ counts/arcsec$^2$/s.  The higher
background in the total image is therefore more than compensated for
by the higher signal--to--noise of the sources.

We checked the detections by eye, removing a spurious double detection
far off--axis, and splitting two blended sources.  We checked also
that point sources included in the brightest extended sources (see \S
3.4) are correctly identified.  Thus, we have 294 sources with S/N
$\geq$ 2.1 from the soft image, and 247 sources from the hard
image. The total number of SExtractor sources is 332.  There are 85
only detected in the soft band, and 38 only detected in the hard band.

\subsection{Wavelet Transform}

To independently check the sources detected with SExtractor, we ran
WAVDETECT (Dobrzycki et al. 1999; Freeman et al. 2001) on the 0.5--7
keV image. A probability threshold of 1 $\times$ 10$^{-6}$, and scales
of 1, 1.414, 2.0, 2.828, 4.0, 5.656, 8.0 pixels were used for source
identification. The combined image of 1 Msec has $\sim$ 1.4 $\times$
10$^6$ pixels; therefore we expect $\sim$ 1.4 false detections per
image for a uniform, static background. However, this is not the case
for our data, where the background is not static and varies across the
field, especially at the edges, between individual observations.

Instead of defining another completeness criterium for the sources
detected with WAVDETECT, and thus a separate catalog, we use this
sample as a complement to the SExtractor catalog.  Therefore we apply
the same S/N$>$2.1 threshold on the list of candidates obtained with
WAVDETECT, removing 23 of them.  Again, the measure of the
signal--to--noise ratio for each source candidate has been done
independently in both the soft and hard images, using the same
extraction regions described in the previous Section.  Finally, we
have 288 sources with S/N $>$ 2.1 from the soft image, and 240 sources
from the hard image. The total number of sources is 318.  There are 78
detected in the soft band only, and 30 detected in the hard band only.

\subsection{Joint Data Set}

We compared the two catalogs from SExtractor and WAVDETECT, using a
matching radius of $2\arcsec$.  We found 304 sources were detected by
both methods.  These jointly detected sources constitute the first
section of the catalog, given in Table~\ref{common}. We also have 28
sources which are detected by SExtractor only, which we list Table 3,
and 14 which are detected by WAVDETECT only, which we list in Table
4. In Figs~\ref{ccsoft} and \ref{cchard} we compare the photon counts
measured with the SExtractor method with the counts measured with
WAVDETECT for each detected source. The comparison has been done both
in the soft and hard band.  We note an excellent agreement between the
independent photometric measurements.  It should also be noted that
all quoted counts in the catalog (Tables 2-4) were determined using
the aperture method described in Section \ref{sextractor} above. The
adopted cut at S/N$>$2.1 corresponds to $\sim 11$($13$) counts for the
faintest sources in the soft(hard) band at the aimpoint.  The net
count rates were obtained by dividing the net counts by the effective
exposure time, which includes the effect of vignetting.

The energy fluxes were computed separately in the soft and hard bands.
We prefer to quote the energy flux in the 2--10 keV band as
extrapolated from the counts in the 2--7 keV bands, in order to make a
direct comparison with previous results from the literature.  To
derive the energy flux from the observed count rate for the soft and
hard bands, we have assumed a conversion factor
appropriate for the measured average spectrum of all the sources
$\Gamma=1.375 \pm 0.015$ (found for all the 1 Msec sources).  After
fixing the local absorption to the Galactic value of 8 $\times$
10$^{19}$ cm$^{-2}$, we have a conversion factor of $(4.6\pm 0.1)
\times 10^{-12}$ from count rate to erg s$^{-1}$ cm$^{-2}$ in the soft
band and $(3.0\pm 0.3) \times 10^{-11}$ in the hard band.  The
uncertainties correspond to $\Gamma=1.4\pm 0.3$.  The error on the
fluxes in the table include only the Poissonian error resulting from
the source and background counts.  With the adopted conversion
factors, the minimum fluxes achieved in the soft and hard band are
$5.5 \times 10^{-17}$ erg s$^{-1}$ cm$^{-2}$ and $4.5 \times 10^{-16}$
erg s$^{-1}$ cm$^{-2}$, respectively.

\subsection{Extended Sources}

The CDFS 1Msec source catalog contains a sizeable fraction of sources
which are resolved by Chandra.  A search for extended sources in the
Msec exposure (Paper III) yielded 18 diffuse sources.  For relatively
rare objects like groups and clusters of galaxies, which are known to
contribute up to $\sim\! 10\%$ of the soft background at $10^{-14}$
erg s$^{-1}$ cm$^{-2}$ (Rosati et al. 1995), the CDFS probes the very
faint end of their X--ray luminosity function.  In order to search for
extended sources associated with hot halos of galaxies, groups and
clusters, we used the soft image and characterized the source extent
with the FWHM of the best fit Gaussian profile.  A higher S/N cut,
$S/N>3$, is required for a robust determination of the source
extent. The likelihood that a source is extended can be estimated as a
function of the off--axis angle by comparing the measured FWHM with the
local PSF width. The latter was empirically derived with the same
profile fitting procedure by using a sample of 346 sources which were
drawn from the combination of the CDFS with two additional deep fields
in the \chandra Archive, MS$1137.5+6625$ (sequence number 800044, 120
ks) and CL0848.6+4453 (Lynx field, 800103, 190 ks). This control
sample was constructed including only sources with $S/N>3$ and
excluding outliers in the extent distribution with a $3\sigma$
clipping procedure.  The best fit of the PSF FWHM as a function of the
off--axis angle is shown in Figure~\ref{FWHM} (top panel) as a solid
line. The coefficients of the parabolic fit are: $(a_0, a_1, a_2) =
(1.152, 0.193, 0.025)$.  This fit was subtracted from each data point
to obtain the FWHM residuals (Figure~\ref{FWHM}, bottom panel). We
then fitted a Gaussian distribution to these residuals in four
different off--axis angle bins (0-3, 3-6, 6-9, 9-12 arcmin) and derived
the $3\sigma$ upper values whose parabolic fit defines our cut off
line for extended sources (dashed line). A catalog of the 18 extended
sources selected in this way is given in Table 5.  We have visually
inspected these sources to make sure that strong variations in the
exposure map do not affect source characterization.

This procedure will generally not work well for very extended low
surface brightness sources for which a simple Gaussian fit is not
appropriate, or which might have been missed by both our standard
detection algorithms.  To search for very diffuse sources we used a
complementary approach.  We ran WAVDETECT with a small characteristic
scale to preferentially select point--like sources.  These objects are
removed from the soft image by replacing the source region with a
simulated background. We then rebinned the resulting image to a larger
pixel size (4 arcsec) and used WAVDETECT to search for significant
sources (probability threshold of 1 $\times$ 10$^{-6}$) with
characteristic scales of 1.0, 1.414, 2.0, 2.828, 4.0, 5.656, 8.0 pixels.
We remind the reader that running WAVDETECT on the original image with
large characteristic scales would produce poor results due to
well-known cross talk between the small and large scales in the
wavelet analysis.  One diffuse source is clearly detected with this
method (\#645 detected by WAVDETECT only). This is a low surface
brightness feature also apparent by visual inspection of the field
(Figure~\ref{image_soft}, lower right). We have added this source to
our extended source catalog. The K--band finding chart
(Figure~\ref{blotch}) shows an asymmetric distribution of X--ray
emission, which is likely associated with an intermediate redshift
poor cluster.

Visual inspection and optical colors indicate that some of them are
groups or isolated early type galaxies at $z<1$.  In several cases
however, the X--ray emission is dominated by a central, hard component
which is likely due to low--level nuclear activity.  Of all the 18
diffuse sources, one has been identified (\#594; see Figure~\ref{cl1})
with a poor cluster at $z=0.72$. The X--ray luminosity is $L_X (0.5-2
\rm{keV}) = 4\times 10^{42}$ erg s$^{-1}$, and the temperature $kT =
1.2^{+0.6}_{-0.3}$ keV (1 sigma errors).  The stacked spectrum of the
three brightest extended sources ($F_X=1-3$ erg s$^{-1}$ cm$^{-2}$ in
the soft band) clearly identified as groups, is well fitted by a
Raymond-Smith model with $kT=1.7^{+0.6}_{-0.4}$ keV, assuming
metallicity 0.3 solar and redshift equal to the spectroscopic or
photometric one. The mean surface brightness of these diffuse sources,
computed as the ratio between the flux and a circular area of 2*FWHM
radius, is as low as $10^{-16}$ erg s$^{-1}$ cm$^{-2}$
$\hbox{arcmin}^{-2}$, i.e.  between 10 and 20 times fainter than the
faintest extended sources discovered by ROSAT.

\section{Optical Identifications \label{opticalIDs}}

Our primary optical imaging was obtained using the FORS1 camera on the
ANTU (UT-1 at VLT) telescope. The $R$ band mosaics from this data
cover $13.6\arcmin\times13.6\arcmin$ to depths between $26$ and $26.7$
(Vega magnitudes).  This data does not cover the full CDFS area and
must be supplemented with other observations.  The ESO Imaging Survey
(EIS) has covered this field to moderate depths in several bands, of
which we present $R$ ($<26.1$ Vega) here (Arnouts et al. 2001; Vandame et
al. 2001).  The EIS data has been obtained using the Wide Field Imager
(WFI) on the ESO-MPG 2.2 meter telescope at La Silla.

The \chandra data itself has a pointing accuracy of roughly 1
$\arcsec$.  Identification of optical counterparts then depends on
registering the X--ray and optical data to the highest possible
accuracy. To do this we required a stable and fixed astrometric
frame. We have chosen to tie all image astrometry to the FORS1 frame
as described below.

\subsection{Positional Accuracy}

The registration of the X--ray to optical coordinates was done by
first cross-correlating the optical and X--ray source catalogs
assuming that any residual offsets were smaller than the search box
size ($10\arcsec$).  After correcting for the bulk offsets, the
correlation was re-run with a smaller box size.  These catalogs were
then used to generate an input file for the IRAF task CCMAP, which
generates an astrometric solution from the correlated image and sky
coordinates. We found shifts in (RA$_R$ - RA$_X$, Dec$_R$ - Dec$_X$) of
(1.1, -0.8) arcseconds from the FORS1 $R$ band to X--ray imaging (see
Figures \ref{fRfit} and \ref{wRfit}).  The measured positional r.m.s.\
is $\sim 0\farcs5$.  We have adopted $1\farcs5$ as our $3\sigma$ error
box.  Due to the strong effect of off--axis angle on the X--ray PSF
and centroid, we scaled the error radius $\delta_r$ with off--axis
angle ($\theta$) and signal--to--noise: $\delta_r = c_{0}*({\textrm
FWHM}^\prime(\theta)/({\textrm S/N}))^{c_{1}}$, where $c_0=0.77$ and
$c_1=0.68$ and FWHM$^\prime$ is the same quadratic as above but is 
re-normalized such that a source at zero off--axis angle would have 
the $1\farcs5$ $3\sigma$ error box.  We are likely over--estimating 
the number of possible counterparts for
some sources.  The resulting optical counterpart candidates presented
in Tables \ref{common}, \ref{sex} and \ref{wav} are sorted by
separation from the X--ray centroid (the optical ids are also marked
with boxes in Figures \ref{maincuts}, \ref{scuts} and \ref{wcuts}).

After applying the shifts and error estimates we find that 85\% of the
sources have a counterpart within the $3\sigma$ error circle. This
percentage is 78\% when calculated with the shallower WFI $R$ band
data.

\subsection{Optically Undetected Sources}

About 15\% of the sources in our sample have no optical counterpart
within the assumed error circle down to the limiting R magnitude. Some
of those sources might have unusually high $f_{x}/f_R$ ratios,
extending the known parameter space of X--ray sources, see
figure~\ref{fxfop} (here $f_R$ is the energy flux in the
$R$--band). We have therefore compared the distribution of the
$f_{x}/f_{R}$ ratios of our X--ray sources without optical
counterparts -- which we assumed have the limiting R magnitude -- with
the analogous distribution for our X--ray sources with optical
counterparts.  As shown in Figure~\ref{fxfrdist}, the distribution of
the $f_{x}/f_{R}$ lower limits of optically undetected sources (dashed
line) is consistent with the high $f_{x}/f_R$ tail of the distribution
of optically detected sources (solid line).  For a concise discussion
of the nature of these sources see \S 5.

\section{Discussion and Conclusions}

We briefly summarize the detection criteria adopted in this Paper.
There are three basic steps: 1) a detection algorithm (either
SExtractor or WAVDETECT) has been run on the full 0.5--7 keV image to
generate a list of source candidates; 2) aperture photometry has been
performed independently in the 0.5--2 keV image and 2--7 keV image at
the position of each source candidate; 3) all those sources with
S/N$>$2.1 either in the soft or in the hard image are considered as
real sources and appear in our catalogs.  This procedure may not reach
the deepest sensitivity achievable with our observations, however, it
is a good compromise achieving low limiting fluxes without introducing
numerous fake sources.  As assessed by extensive simulations described
in \citet{tozzi2001}, the fraction of fake sources to be expected in
our sample is less than 10 (or $\sim 3$\%) with this method.  The
faintest sources in our samples have approximately 10 net counts.
Given the exposure time and the adopted conversion factors the
limiting fluxes achieved in the 0.5--2 keV and in the 2--10 keV band
are $5.5\times10^{-17}$ erg s$^{-1}$ cm$^{-2}$ and $4.5\times10^{-16}$
erg s$^{-1}$ cm$^{-2}$, respectively.

These limiting fluxes are about 20 and 200 times lower than that
previous X--ray missions in the soft and hard bands respectively.  The
catalog of sources presented in this Paper allows us to investigate
the nature of X--ray faint sources with 
unprecedented sensitivity, together with their optical properties.
Most of the sources in our sample fall within the X--ray to optical
flux ratio range typical of AGN as determined from the EMSS sources
(Stocke et al. 1991).  In addition a significant population of X--ray
faint/optically bright sources is observed, with $f_x/f_R\leq 0.1$, lower than
those typical of AGNs. Most of these sources are detected
in the soft band only (circles in Figure ~\ref{fxfop}).  This
population of sources has been already partially identified as nearby,
bright normal galaxies by
\citet{tozzi2001,hornschemeier2001,barger2001}.  Only a few sources
seem to have $f_{x}/f_{R}$ values higher than usual, which may be
obscured AGN at very high redshift.  This population of optically
faint sources ($R>25$) is consistent with a mix of obscured AGN at
$z=1-3$ and evolved, high--z galaxies (see Alexander et al. 2001;
Cowie et al. 2001).  The nature of these sources will be investigated
in greater detail in following papers.

The data presented here constitute (along with the \chandra Deep Field
North) the deepest X--ray exposure ever taken. As such, it is a unique
dataset for current and future research.  A proposal to re--observe
this field with \chandra will await detailed analysis of the current
dataset.  However, a SIRTF Legacy program will be observing this field
with both MIPS and IRAC during their own deep survey. Also, a 500ks
observation with XMM has been planned in the CDFS this year. The high
XMM throughput, combined with the arcsec \chandra resolution, will
allow high quality X--ray spectra and secure optical identifications
to be obtained for most of the X--ray sources in the CDFS. The
intensive optical and infrared coverage is planned to continue and to
be vigorously pursued at ESO. Deep VLA radio observations are being
analyzed.

\acknowledgements

We thank Dr. Harvey Tananbaum for making available 500 kiloseconds of
Director Discretionary Time which doubled our original guaranteed time
on this field, thus permitting us to achieve the quoted sensitivity.
We thank the entire \chandra Team for the high degree of support we
have received in carrying out our observing program.  In particular,
we wish to thank Antonella Fruscione for her constant help in the use
of the CXC software.  Finally, we thank the anonymous referee for a
detailed report that considerably improved the presentation of the
results.  R. Giacconi and C. Norman gratefully acknowledge support
under NASA grant NAG-8-1527 and NAG-8-1133.

\begin{deluxetable}{llllll}
\tabletypesize{\scriptsize} \tablewidth{0pt} \tablecaption{\chandra
observations of CDFS \label{ObsTab}} \tablehead{ \colhead{Obs. ID} &
\colhead{Obs. Date} & \colhead{Exposure Time (ks) \tablenotemark{a}} &
\colhead{Roll Angle} & \colhead{Aim Point} & \colhead{Count Rate\tablenotemark{b}}} 
\startdata 
1431-0 & 1999 Oct 15 & 24.983 & 47.28 & 3 32 29.4 -27 48 21.8 & 0.980\\  
1431-1 & 1999 Nov 23 & 92.807 & 47.28 & 3 32 29.4 -27 48 21.8& 0.7846\\ 
441 & 2000 May 27 & 55.727 & 166.73 & 3 32 26.8 -27 48 17.4& 0.6812\\ 
582 & 2000 Jun 03 & 129.869 & 162.93 &3 32 26.8 -27 48 16.4& 0.6984\\ 
2406 & 2000 Dec 10 & 29.564 & 332.18 &3 32 28.4 -27 48 39.3& 0.668\\ 
2405 & 2000 Dec 11 & 59.363 & 331.81 &3 32 29.0 -27 48 46.4& 0.670\\ 
2312 & 2000 Dec 13 & 123.212 & 329.92 &3 32 28.4 -27 48 39.8& 0.6528\\ 
1672 & 2000 Dec 16 & 94.564 & 326.90 &3 32 28.9 -27 48 47.5& 0.6589\\ 
2409 & 2000 Dec 19  & 68.719 & 319.21 &3 32 28.2 -27 48 41.8& 0.6570\\ 
2313 & 2000 Dec 21 & 129937 & 319.21 &3 32 28.2 -27 48 41.9& 0.65487\\ 
2239 & 2000 Dec 23 & 130.250 & 319.21 &3 32 28.2 -27 48 41.8& 0.674\\ 
\enddata

\tablenotetext{a}{effective exposure time after cleaning bad aspect
interval and high background intervals (about 8800 s are lost due to
high background)} 
\tablenotetext{b}{Count rates on four ACIS-I CCD (0.5--7 keV)}
\end{deluxetable}

\clearpage

\begin{deluxetable}{r@{}r@{~~}r@{}r@{}r@{~~~}r@{~}r@{}r@{}r@{}r@{}r@{~~}r@{~~~}r@{}r@{~}r@{~~~~}r}
\rotate
\tabletypesize{\tiny}
\tablewidth{0pt}
\tablecaption{Main Catalog \label{common}}
\tablehead{
\colhead{ID\tablenotemark{a}} & \colhead{XID\tablenotemark{b}} & \colhead{CXO CDFS\tablenotemark{c}} & \colhead{RA (J2000)\tablenotemark{d}} & \colhead{Dec (J2000)\tablenotemark{e}} &
\colhead{Soft Cts.\tablenotemark{f}} & \colhead{Hard Cts.\tablenotemark{g}} & \colhead{Exp Time\tablenotemark{h}} & \colhead{Soft Flux\tablenotemark{i}} & \colhead{Error\tablenotemark{j}} & \colhead{Hard Flux\tablenotemark{k}} & \colhead{Error\tablenotemark{l}} &
\colhead{HR\tablenotemark{m}} & \colhead{Pos. Err.\tablenotemark{n}} & \colhead{Sep.\tablenotemark{o}} & \colhead{$R$\tablenotemark{p}} }
\startdata
1 & 576 & J033144.3-274927 & 03 31 44.28  & -27 49 27.19  & $  27.00\pm11.5$ & \nodata & 438.4(433.7) &2.29e-16 & 9.77e-17 & $<$1.35e-15 & \nodata & $ -1.00\pm0.00$ &   2.44 & \nodata & $>26.10$ W\\
2 & 548 & J033144.7-275158 & 03 31 44.73  & -27 51 58.21  & $  88.80\pm15.3$ & $  35.30\pm17.4$ & 584.7(575.8) &6.35e-16 & 1.09e-16 & 1.62e-15 & 7.96e-16 & $ -0.42\pm0.21$ &   1.84 & \nodata & $>26.10$ W\\
3 & 591 & J033145.0-275139 & 03 31 44.96  & -27 51 39.71  & $  60.20\pm14.5$ & \nodata & 620.2(611.8) &4.06e-16 & 9.78e-17 & $<$1.31e-15 & \nodata & $ -1.00\pm0.00$ &   1.99 & \nodata & $>26.10$ W\\
4 & 209 & J033147.3-275314 & 03 31 47.33  & -27 53 14.96  & $ 320.70\pm19.7$ & $ 159.10\pm17.8$ & 214.1(210.7) &3.75e-15 & 2.31e-16 & 1.19e-14 & 1.33e-15 & $ -0.33\pm0.06$ &   1.60 &   0.49 & $ 25.02 \pm   0.28$ W\\
5 & 238 & J033148.0-275046 & 03 31 48.02  & -27 50 46.82  & $ 315.90\pm20.5$ & $ 115.80\pm17.1$ & 647.3(644.7) &2.10e-15 & 1.36e-16 & 4.88e-15 & 7.21e-16 & $ -0.46\pm0.06$ &   1.60 &   0.64 & $ 23.27 \pm   0.09$ W\\
6 & 569 & J033148.1-274802 & 03 31 48.12  & -27 48 2.84  & $  74.70\pm11.9$ & \nodata & 575.1(573.8) &5.60e-16 & 8.92e-17 & $<$1.09e-15 & \nodata & $ -1.00\pm0.00$ &   1.79 &   0.29 & $ 24.49 \pm   0.18$ W\\
7 & 604 & J033148.6-274715 & 03 31 48.65  & -27 47 15.58  & \nodata & $  47.40\pm11.8$ & 397.6(396.6) &$<$1.53e-16 & \nodata & 2.91e-15 & 7.23e-16 & $  1.00\pm0.00$ &   1.99 &   0.73 & $ 26.56 \pm   0.48$ W\\
8 & 225 & J033149.5-274634 & 03 31 49.50  & -27 46 34.75  & $  97.40\pm12$ & $  51.10\pm11.5$ & 256.9(256.3) &1.04e-15 & 1.28e-16 & 3.45e-15 & 7.77e-16 & $ -0.31\pm0.12$ &   1.71 &   0.56 & $ 22.31 \pm   0.04$ W\\
\enddata
\tablenotetext{a}{Optical cutout ID number, in order of RA}
\tablenotetext{b}{Unique Detection ID}
\tablenotetext{c}{IAU Registered Name, based on original (not offset) X--ray coordinates}
\tablenotetext{d}{Right Ascension, offset from {\it Chandra} by $-1.2\arcsec$}
\tablenotetext{e}{Declination, offset from {\it Chandra} by $0.8\arcsec$}
\tablenotetext{f}{Net Counts in soft (0.5 - 2 keV) band}
\tablenotetext{g}{Net Counts in hard (2 - 10 keV) band}
\tablenotetext{h}{Effective exposure time in the soft(hard) band, in kiloseconds}
\tablenotetext{i}{Flux in soft (0.5 - 2 keV) band, c.g.s. units}
\tablenotetext{j}{Error on soft flux}
\tablenotetext{k}{Flux in hard (2 - 10 keV) band, c.g.s. units}
\tablenotetext{l}{Error on hard flux}
\tablenotetext{m}{Hardness Ratio, defined as (H - S)/(H + S) where H and S are the net counts in the hard and soft bands respectively}
\tablenotetext{n}{Error on the X-ray source position as calculated in the text, in arcseconds}
\tablenotetext{o}{Separation of the candidate optical counterpart, in arcseconds}
\tablenotetext{p}{Vega $R$ magnitude for the optical counterpart in either the FORS (F) or WFI (W) filter}
\end{deluxetable}

\begin{deluxetable}{r@{}r@{~~}r@{}r@{}r@{~~~}r@{~}r@{}r@{}r@{}r@{}r@{~~}r@{~~~}r@{}r@{~}r@{~~~~}r}
\rotate
\tabletypesize{\tiny}
\tablewidth{0pt}
\tablecaption{SExtractor Only \label{sex}}
\tablehead{
\colhead{ID\tablenotemark{a}} & \colhead{XID\tablenotemark{b}} & \colhead{CXO CDFS\tablenotemark{c}} & \colhead{RA (J2000)\tablenotemark{d}} & \colhead{Dec (J2000)\tablenotemark{e}} &
\colhead{Soft Cts.\tablenotemark{f}} & \colhead{Hard Cts.\tablenotemark{g}} & \colhead{Exp Time\tablenotemark{h}} & \colhead{Soft Flux\tablenotemark{i}} & \colhead{Error\tablenotemark{j}} & \colhead{Hard Flux\tablenotemark{k}} & \colhead{Error\tablenotemark{l}} &
\colhead{HR\tablenotemark{m}} & \colhead{Pos. Err.\tablenotemark{n}} & \colhead{Sep.\tablenotemark{o}} & \colhead{$R$\tablenotemark{p}} }
\startdata
1 & 636 & J033150.5-275043 & 03 31 50.45  & -27 50 43.62  & \nodata & $  33.60\pm13.2$ & 657.5(658.4) &$<$1.23e-16 & \nodata & 1.39e-15 & 5.45e-16 & $  1.00\pm0.00$ &   2.24 &   1.08 & $ 25.50 \pm   0.30$ W\\
2 & 633 & J033150.5-275213 & 03 31 50.52  & -27 52 13.33  & \nodata & $  34.30\pm14.9$ & 732.0(729.8) &$<$1.34e-16 & \nodata & 1.38e-15 & 5.99e-16 & $  1.00\pm0.00$ &   2.34 & \nodata & $>26.10$ W\\
3 & 619 & J033155.6-275403 & 03 31 55.62  & -27 54 3.49  & $  23.60\pm11$ & \nodata & 757.9(754.3) &1.45e-16 & 6.75e-17 & $<$9.97e-16 & \nodata & $ -1.00\pm0.00$ &   2.48 & \nodata & $>26.10$ W\\
4 & 625 & J033200.9-274758 & 03 32 0.88  & -27 47 58.88  & $  14.70\pm6.2$ & \nodata & 876.2(897.8) &8.02e-17 & 3.38e-17 & $<$5.10e-16 & \nodata & $ -1.00\pm0.00$ &   2.20 &   1.86 & $ 21.68 \pm   0.02$ F\\
5 & 615 & J033201.3-275052 & 03 32 1.34  & -27 50 52.26  & $  14.80\pm6.5$ & $  17.80\pm7.8$ & 799.3(814.7) &8.86e-17 & 3.89e-17 & 6.59e-16 & 2.89e-16 & $  0.08\pm0.31$ &   2.25 &   0.69 & $ 23.41 \pm   0.09$ F\\
6 & 626 & J033209.5-274758 & 03 32 9.47  & -27 47 58.92  & $  17.20\pm6.1$ & \nodata & 877.1(908.1) &9.38e-17 & 3.33e-17 & $<$4.64e-16 & \nodata & $ -1.00\pm0.00$ &   2.00 &   1.58 & $ 25.35 \pm   0.17$ F\\
7 & 178 & J033213.9-275034 & 03 32 13.89  & -27 50 34.84  & $  15.90\pm5.9$ & \nodata & 904.3(930.5) &8.41e-17 & 3.12e-17 & $<$4.29e-16 & \nodata & $ -1.00\pm0.00$ &   2.00 & \nodata & $> 26.26$ F\\
8 & 631 & J033215.2-274335 & 03 32 15.19  & -27 43 35.94  & $  19.40\pm6.3$ & \nodata & 878.8(892.5) &1.06e-16 & 3.43e-17 & $<$4.84e-16 & \nodata & $ -1.00\pm0.00$ &   2.01 &   0.75 & $ 24.35 \pm   0.20$ F\\
\enddata
\tablenotetext{a}{Optical cutout ID number, in order of RA}
\tablenotetext{b}{Unique Detection ID}
\tablenotetext{c}{IAU Registered Name, based on original (not offset) X--ray coordinates}
\tablenotetext{d}{Right Ascension, offset from {\it Chandra} by $-1.2\arcsec$}
\tablenotetext{e}{Declination, offset from {\it Chandra} by $0.8\arcsec$}
\tablenotetext{f}{Net Counts in soft (0.5 - 2 keV) band}
\tablenotetext{g}{Net Counts in hard (2 - 10 keV) band}
\tablenotetext{h}{Effective exposure time in the soft(hard) band, in kiloseconds}
\tablenotetext{i}{Flux in soft (0.5 - 2 keV) band, c.g.s. units}
\tablenotetext{j}{Error on soft flux}
\tablenotetext{k}{Flux in hard (2 - 10 keV) band, c.g.s. units}
\tablenotetext{l}{Error on hard flux}
\tablenotetext{m}{Hardness Ratio, defined as (H - S)/(H + S) where H and S are the net counts in the hard and soft bands respectively}
\tablenotetext{n}{Error on the X-ray source position as calculated in the text, in arcseconds}
\tablenotetext{o}{Separation of the candidate optical counterpart, in arcseconds}
\tablenotetext{p}{Vega $R$ magnitude for the optical counterpart in either the FORS (F) or WFI (W) filter}
\end{deluxetable}

\begin{deluxetable}{r@{}r@{~~}r@{}r@{}r@{~~~}r@{~}r@{}r@{}r@{}r@{}r@{~~}r@{~~~}r@{}r@{~}r@{~~~~}r}
\rotate
\tabletypesize{\tiny}
\tablewidth{0pt}
\tablecaption{Wavdetect Only \label{wav}}
\tablehead{
\colhead{ID\tablenotemark{a}} & \colhead{XID\tablenotemark{b}} & \colhead{CXO CDFS\tablenotemark{c}} & \colhead{RA (J2000)\tablenotemark{d}} & \colhead{Dec (J2000)\tablenotemark{e}} &
\colhead{Soft Cts.\tablenotemark{f}} & \colhead{Hard Cts.\tablenotemark{g}} & \colhead{Exp Time\tablenotemark{h}} & \colhead{Soft Flux\tablenotemark{i}} & \colhead{Error\tablenotemark{j}} & \colhead{Hard Flux\tablenotemark{k}} & \colhead{Error\tablenotemark{l}} &
\colhead{HR\tablenotemark{m}} & \colhead{Pos. Err.\tablenotemark{n}} & \colhead{Sep.\tablenotemark{o}} & \colhead{$R$\tablenotemark{p}} }
\startdata
1 & 645 & J033150.0-274941 & 03 31 49.95  & -27 49 41.74  & $  66.50\pm13$ & \nodata & 620.1(621.2) &4.62e-16 & 9.04e-17 & $<$1.00e-15 & \nodata & $ -1.00\pm0.00$ &   1.86 &   1.83 & $ 21.00 \pm   0.02$ W\\
2 & 643 & J033156.4-275257 & 03 31 56.43  & -27 52 57.00  & $  33.90\pm10$ & $  28.90\pm12.5$ & 821.3(824.4) &1.97e-16 & 5.82e-17 & 1.06e-15 & 4.57e-16 & $ -0.08\pm0.26$ &   2.05 & \nodata & $>26.10$ W\\
3 & 124 & J033202.6-274525 & 03 32 2.65  & -27 45 25.20  & $  14.80\pm6.8$ & \nodata & 816.4(827.9) &8.67e-17 & 3.98e-17 & $<$6.08e-16 & \nodata & $ -1.00\pm0.00$ &   2.30 &   0.49 & $ 25.49 \pm   0.27$ F\\&&&&&&&&&&&&&&   1.32 & $ 25.29 \pm   0.26$ F\\
&&&&&&&&&&&&&&   2.20 & $ 24.94 \pm   0.23$ F\\
4 & 642 & J033215.3-274159 & 03 32 15.35  & -27 41 59.46  & $  15.40\pm7.5$ & $  25.60\pm10$ & 838.8(841.7) &8.78e-17 & 4.28e-17 & 9.17e-16 & 3.58e-16 & $  0.25\pm0.29$ &   2.40 &   1.07 & $ 24.54 \pm   0.25$ F\\
5 & 649 & J033224.9-273851 & 03 32 24.85  & -27 38 51.54  & $  21.80\pm10.3$ & \nodata & 391.7(386.2) &2.15e-16 & 1.01e-16 & $<$1.52e-15 & \nodata & $ -1.00\pm0.00$ &   2.57 & \nodata & $> 25.72$ F\\
6 & 651 & J033228.6-275809 & 03 32 28.56  & -27 58 9.95  & $  44.60\pm10.9$ & \nodata & 361.3(353.8) &4.16e-16 & 1.02e-16 & $<$1.37e-15 & \nodata & $ -1.00\pm0.00$ &   2.00 &   1.29 & $ 18.97 \pm   0.01$ W\\
\enddata
\tablenotetext{a}{Optical cutout ID number, in order of RA}
\tablenotetext{b}{Unique Detection ID}
\tablenotetext{c}{IAU Registered Name, based on original (not offset) X--ray coordinates}
\tablenotetext{d}{Right Ascension, offset from {\it Chandra} by $-1.2\arcsec$}
\tablenotetext{e}{Declination, offset from {\it Chandra} by $0.8\arcsec$}
\tablenotetext{f}{Net Counts in soft (0.5 - 2 keV) band}
\tablenotetext{g}{Net Counts in hard (2 - 10 keV) band}
\tablenotetext{h}{Effective exposure time in the soft(hard) band, in kiloseconds}
\tablenotetext{i}{Flux in soft (0.5 - 2 keV) band, c.g.s. units}
\tablenotetext{j}{Error on soft flux}
\tablenotetext{k}{Flux in hard (2 - 10 keV) band, c.g.s. units}
\tablenotetext{l}{Error on hard flux}
\tablenotetext{m}{Hardness Ratio, defined as (H - S)/(H + S) where H and S are the net counts in the hard and soft bands respectively}
\tablenotetext{n}{Error on the X-ray source position as calculated in the text, in arcseconds}
\tablenotetext{o}{Separation of the candidate optical counterpart, in arcseconds}
\tablenotetext{p}{Vega $R$ magnitude for the optical counterpart in either the FORS (F) or WFI (W) filter}
\end{deluxetable}

\begin{deluxetable}{rrrrrrrr}
\rotate
\tabletypesize{\tiny}
\tablewidth{0pt}
\tablecaption{Extended Sources \label{extended}}
\tablehead{
\colhead{XID} & \colhead{RA (J2000)} & \colhead{Dec (J2000)} & 
\colhead{Soft Cts} & \colhead{Soft Flux} & \colhead{S/N$_{\textrm soft}$} 
& \colhead{FWHM($\arcsec$)} & \colhead{Off-axis Angle($\arcmin$)}}
\startdata
   37 &     3 32 32.21 &    -27 41 56.22&      89.8 & 5.03e-16 & 8.0 &       5.6 &      6.77 \\
  116 &     3 32 30.07 & -27 44 5.35 &      93.0 & 5.29e-16 & 8.8 &       3.9 &      4.58 \\
  132 &     3 32 44.09 & -27 54 56.12  &      40.9 & 2.19e-16 & 4.8 &       6.9 &      7.17 \\
  138 &     3 32 50.11 & -27 41 35.30  &      51.8 & 9.13e-16 & 5.9 &       7.3 &      8.54 \\
  147 &     3 32 46.43 & -27 46 32.74  &      25.9 & 1.49e-16 & 4.0 &       4.1 &      4.50 \\
  240 &     3 32 59.20 & -27 51 40.43  &      43.8 & 2.46e-16 & 4.9 &       7.4 &      7.43 \\
  249 &     3 32 19.49 & -27 54 6.55   &      27.4 & 1.46e-16 & 3.7 &       8.2 &      5.80 \\
  507 &     3 33 0.12  & -27 49 25.50   &      23.3 & 1.43e-16 & 3.0 &       5.7 &      7.04 \\
\enddata
\end{deluxetable}

\clearpage

\begin{figure}
\caption{Soft(0.5--2 keV) band image of 942 ks exposure of the
CDFS. The image is smoothed with a Gaussian with $\sigma=1\arcsec$.}
\label{image_soft}
\end{figure}

\begin{figure}
\caption{Hard (2--7 keV) band image of 942 ks exposure of the
CDFS. The image is smoothed with a Gaussian with $\sigma=1\arcsec$.}
\label{image_hard}
\end{figure}

\begin{figure}
\caption{Exposure map of the combined 11 exposures of the CDFS,
computed for an energy of 1.5 keV.  }
\label{expmap}
\end{figure}

\begin{figure}
\plotone{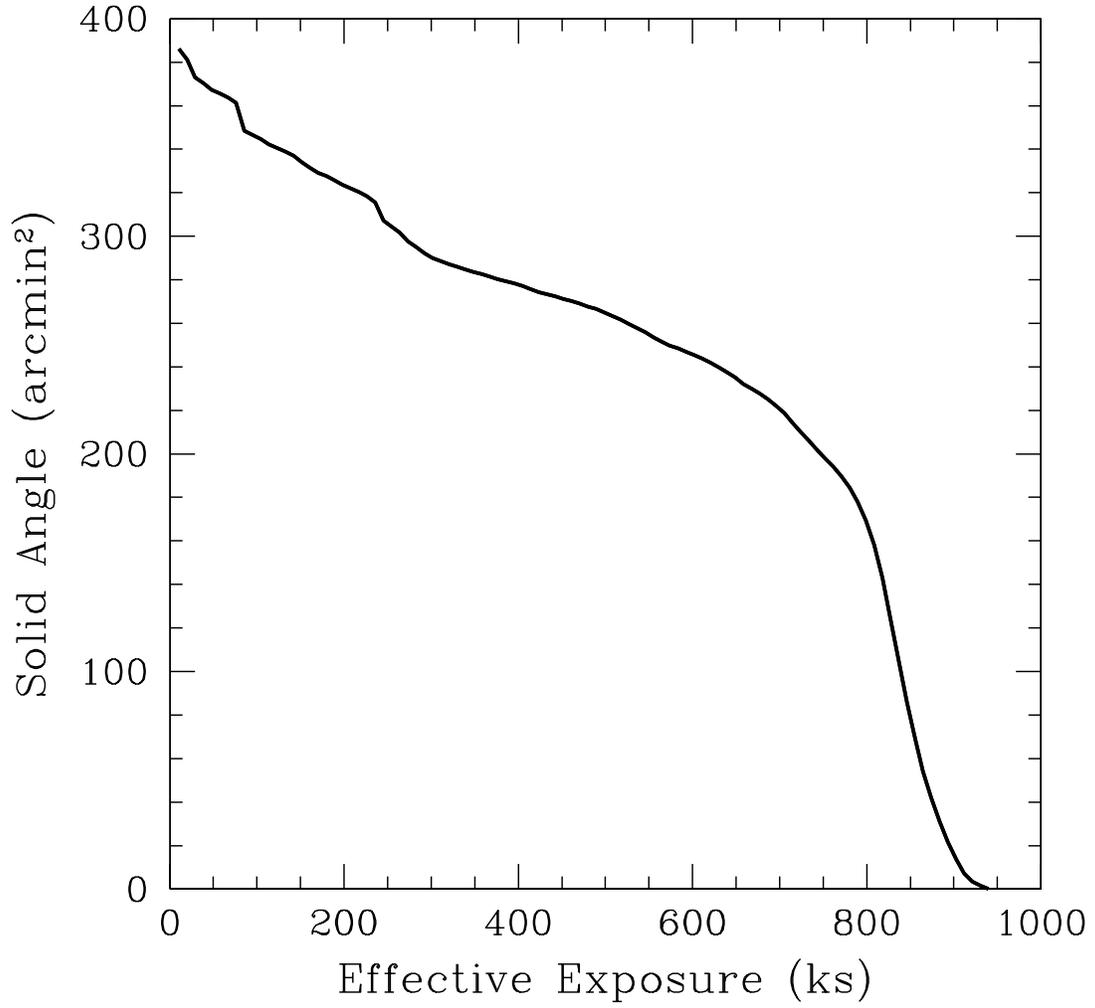}
\caption{Solid angle as a function of the effective exposure time in
the soft band.}
\label{eff_time}
\end{figure}

\begin{figure}
\plotone{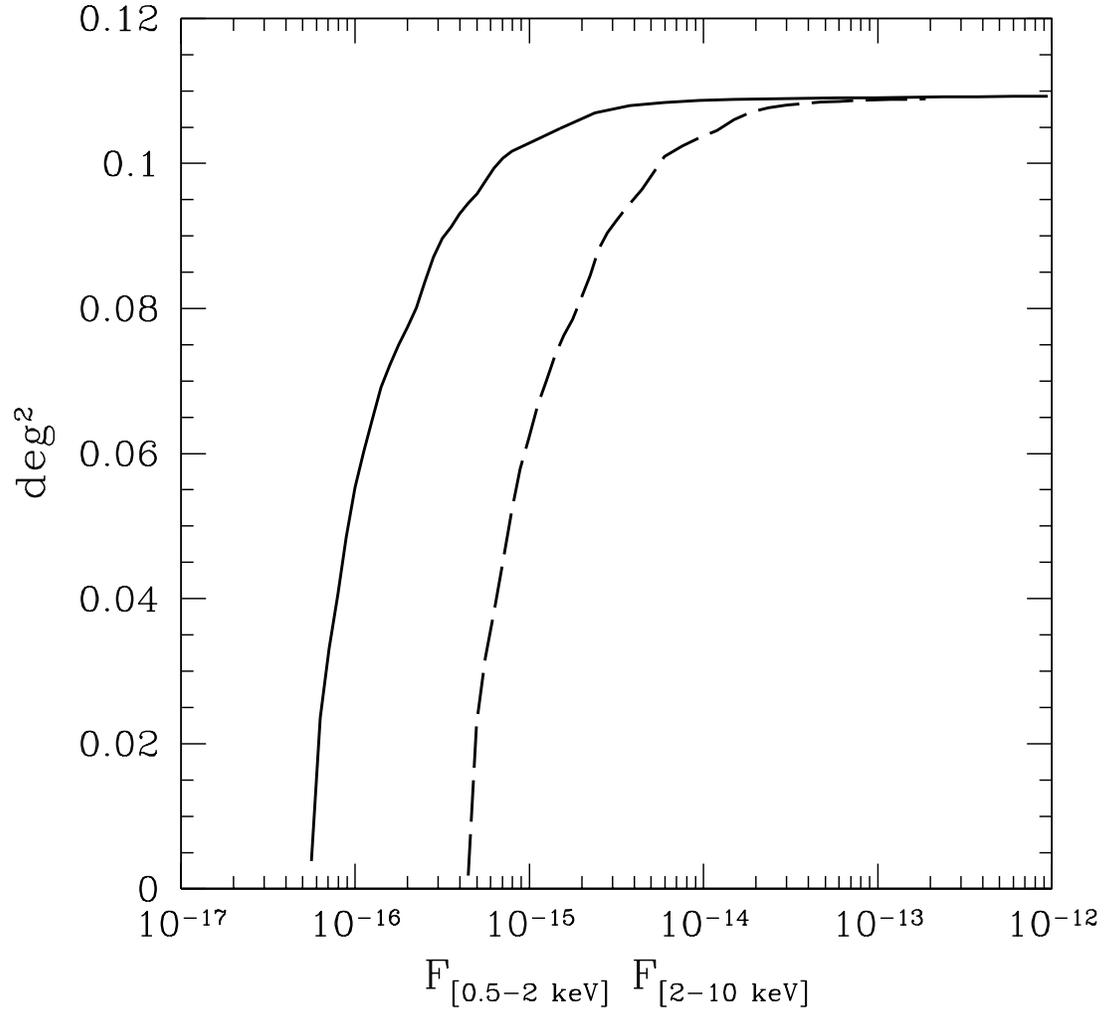}
\caption{The sky coverage (area covered vs. flux limit) for the soft
image (solid line) and the hard image (dashed line).  The sky coverage
is plotted against the 2--10 keV energy flux which is actually used in
the catalog.  }
\label{skycov}
\end{figure}

\clearpage

\begin{figure}
\plotone{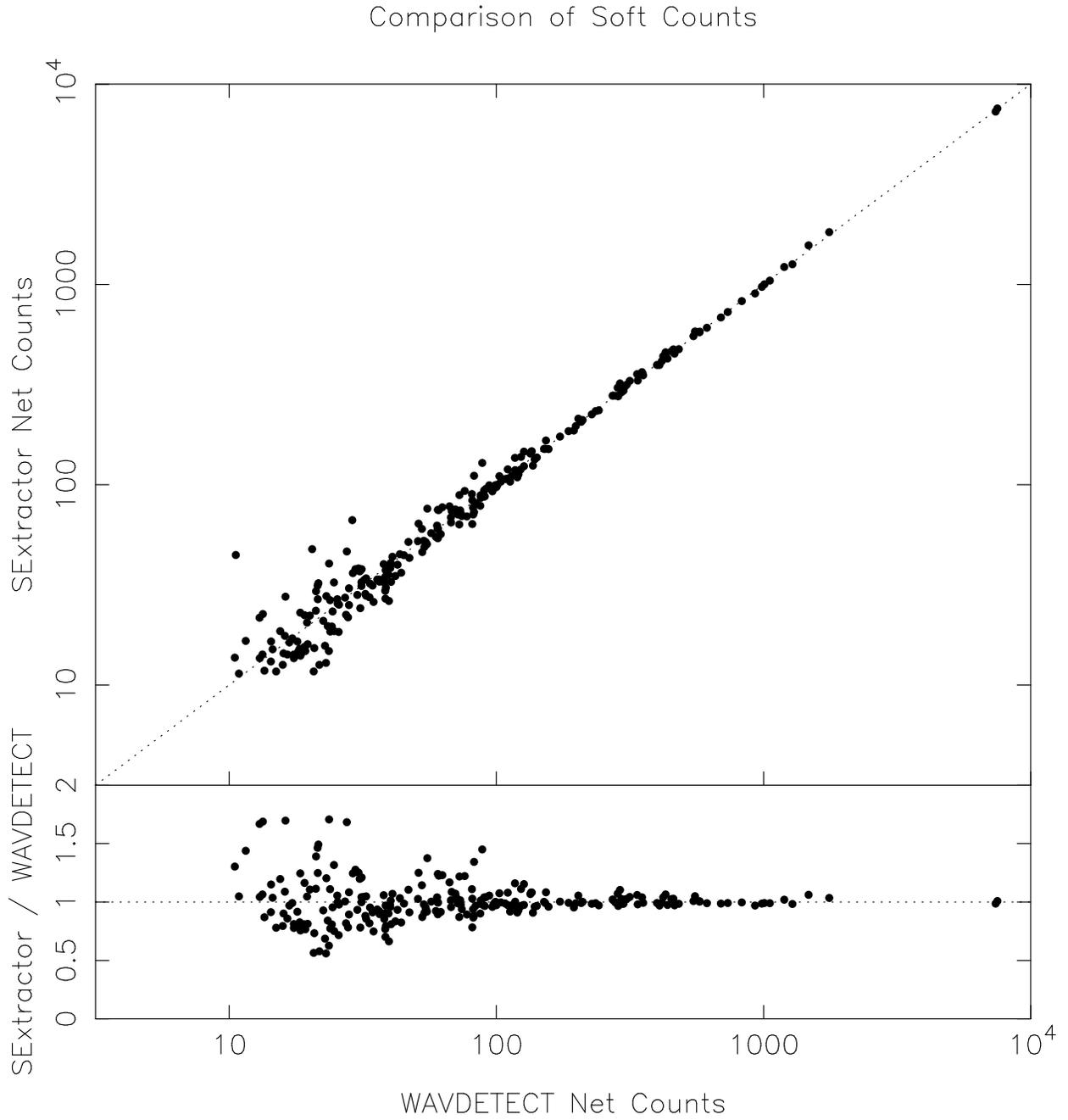}
\caption{Comparison between the photon counts obtained with the
SExtractor method and those obtained with the WAVDETECT method for the
sources detected in the 0.5-2 keV band.}
\label{ccsoft}
\end{figure}

\begin{figure}
\plotone{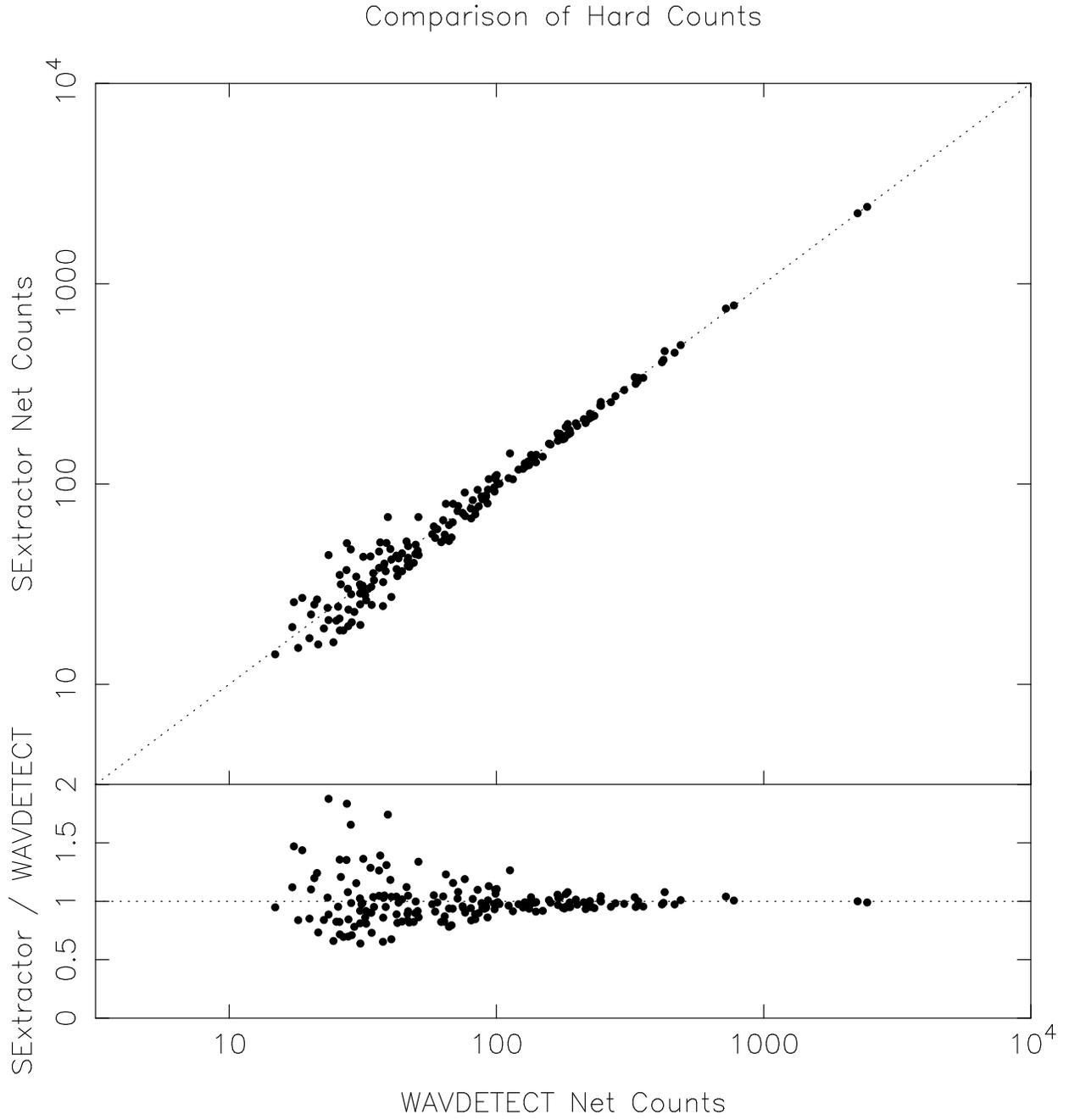}
\caption{The same as in Fig.~\ref{ccsoft} but for the sources detected
in the 2--7 keV band}
\label{cchard}
\end{figure}

\begin{figure}
\plotone{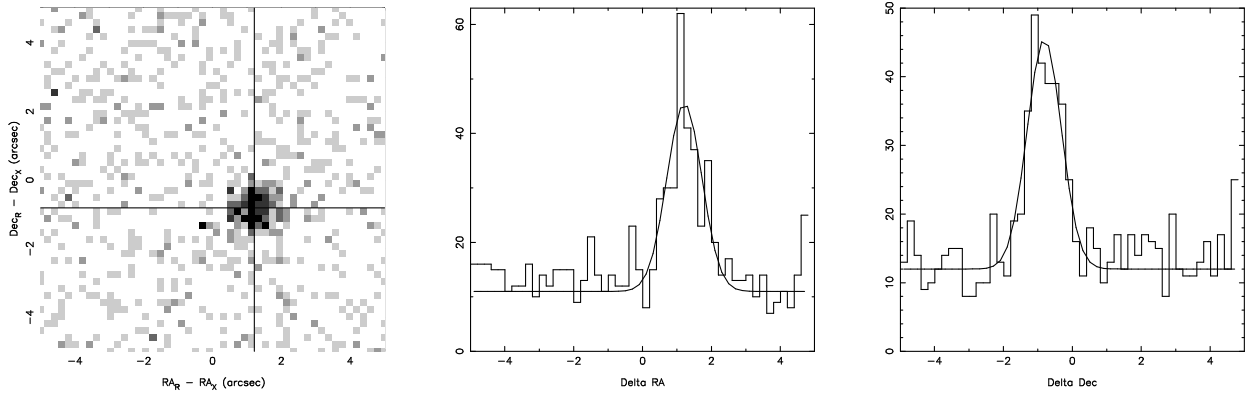}
\caption{Coordinate offsets between \chandra and FORS R.  The 2D
histogram shows the correlation peak between the two catalogs.  Each
1D distribution was fit with a Gaussian to determine both the mean and
scatter of the coordinate offsets. \label{fRfit}}
\end{figure}

\begin{figure}
\plotone{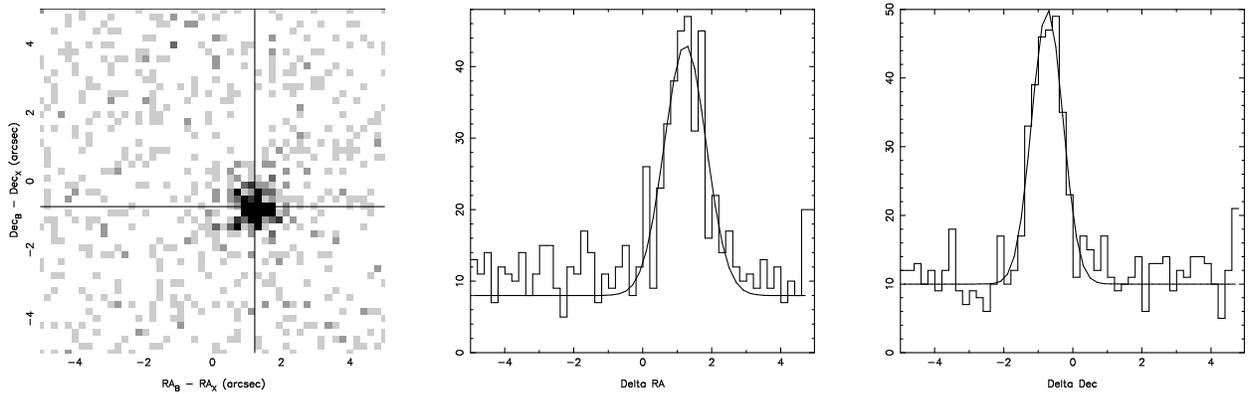}
\caption{Coordinate offsets between \chandra and WFI R.  Identical to
Figure~\ref{fRfit}. \label{wRfit}}
\end{figure}

\clearpage

\begin{figure}
\caption{Selection of extended sources in the FWHM vs off--axis angle
plane. Top panel: Crosses are sources in CDFS (with S/N $> 3$ in the
soft band), diamonds are sources drawn from two additional archival
deep Chandra fields. The solid line is a parabolic best fit to the
PSF, dashed and dot-dashed lines are the $3\sigma$ and $2\sigma$
limits of the FWHM distribution respectively (see text).  Filled dots
are sources which are likely to be extended.  Bottom panel:
distribution of residuals for the CDFS sources only, after subtracting
the PSF fit model.}
\label{FWHM}
\end{figure}

\begin{figure}
\caption{K--band image of source XID 645, one of the extended low
surface--brightness sources in the CDFS.  The image is from the EIS
data of the CDFS \citep{vandame2001} with overlaid Chandra contours
corresponding to $[2,3,5,7,15]\sigma$ above the local background. The
soft band image smoothed with a $\sigma=5\arcsec$ Gaussian was used.}
\label{blotch}
\end{figure}

\begin{figure}
\caption{$R$--band FORS mosaic image of the brightest extended source
in the CDFS (XID 594, identified as a poor cluster at $z=0.72$) with
overlaid \chandra contours corresponding to
$[2,4,5,7,10,15,200]\sigma$ above the local background. The soft band
image smoothed with a $\sigma=2.5\arcsec$ Gaussian was used (upper
left inset).}
\label{cl1}
\end{figure}

\clearpage

\begin{figure}
\caption{$f_{X}$ vs. $f_{R}$ for the CDFS sample.  We computed the
energy flux $f_X$ in the 0.5--10 keV band.  The circles and squares
indicate sources detected in only the soft or hard band respectively.
The arrows indicate lower limit in the magnitude of optically
undetected sources.  The differences in these lower limits reflects
the different depth reached by the FORS images, from where the R
magnitueds were derived.  The three dashed lines are lines of
constant $f_{X}/f_{R}$, with values left to right of 0.1, 1 and
16.\label{fxfop}}
\end{figure}

\clearpage 

\begin{figure}
\plotone{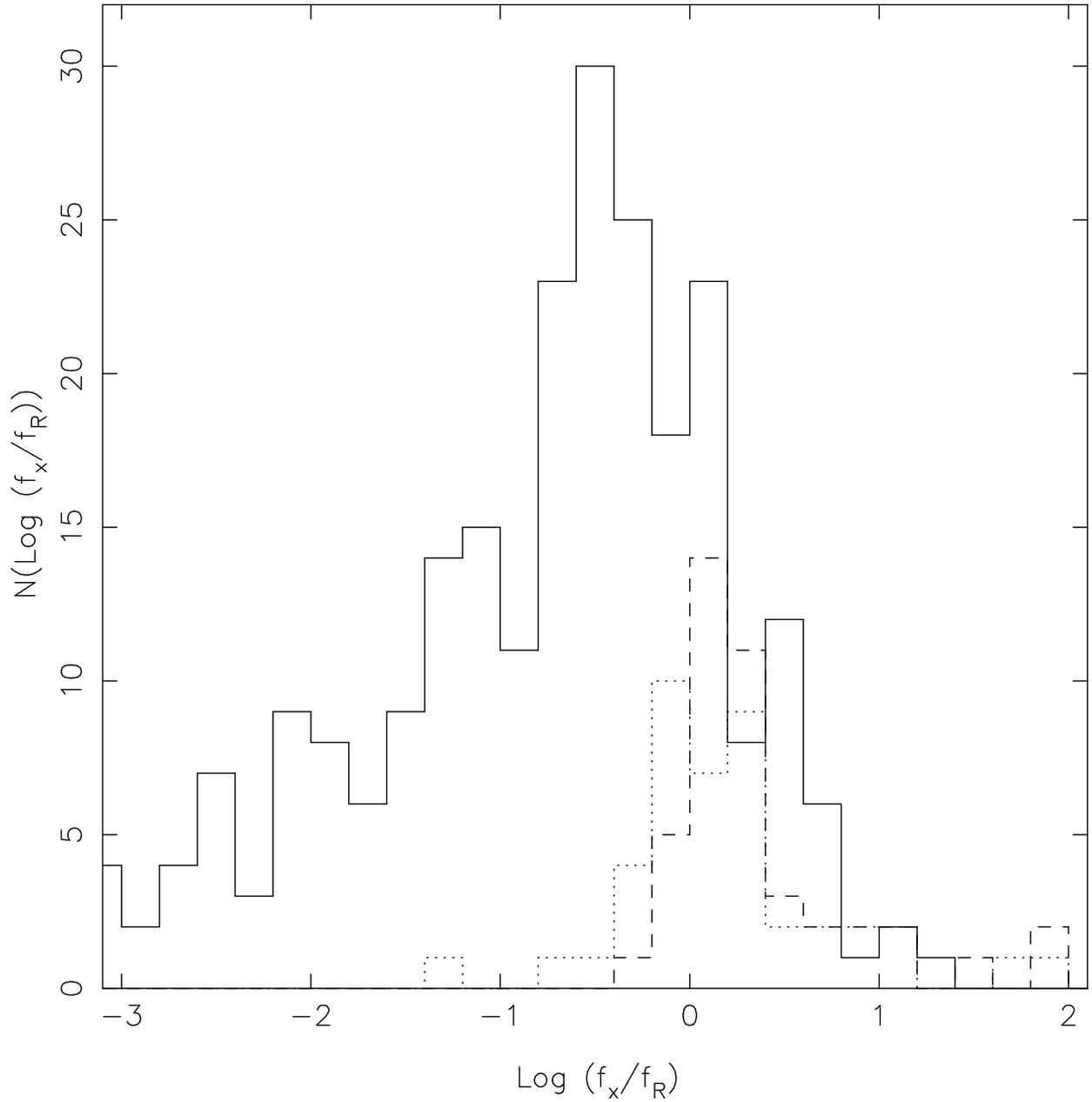}
\caption{The distribution of Log $f_{X}/f_{R}$ for the current sample.
The solid line histogram is for the CDFS data, the dotted line is the
{\it ROSAT} Lockman Hole data \citep*{schmidt1998} and the dashed line
is upper limits in the CDFS/FORS1 dataset for the optically undetected
counterparts (arrows in Figure \ref{fxfop}).  }
\label{fxfrdist}
\end{figure}

\clearpage

\begin{figure}
\caption{Optical cutouts for the Main Catalog.  ID number (XID in
parenthesis) and optical band is printed in the upper left corner of
each image.  The energy flux in the soft band is in the lower left,
while the flux in the hard 2--10 keV band is in the lower right.  The
iso-intensity contours are at 3, 5, 10, 20 and 100 sigma above the
local background.  The box indicates an optical counterpart candidate.
The 3$\sigma$ positional error (after the relation in \S 4.1) is
indicated by the circle in the center.  Each cutout is 20 arcsec wide.
\label{maincuts}}
\end{figure}

\begin{figure}
\end{figure}

\clearpage

\begin{figure}
\end{figure}

\clearpage

\begin{figure}
\end{figure}

\clearpage

\begin{figure}
\end{figure}

\clearpage

\begin{figure}
\end{figure}

\clearpage

\begin{figure}
\end{figure}

\clearpage

\begin{figure}
\end{figure}

\clearpage

\begin{figure}
\end{figure}

\clearpage

\begin{figure}
\end{figure}

\clearpage

\begin{figure}
\end{figure}

\clearpage

\begin{figure}
\end{figure}

\clearpage

\begin{figure}
\end{figure}

\clearpage

\begin{figure}
\end{figure}

\clearpage

\begin{figure}
\end{figure}

\clearpage

\begin{figure}
\end{figure}

\clearpage

\begin{figure}
\end{figure}

\clearpage

\begin{figure}
\end{figure}

\clearpage

\begin{figure}
\end{figure}

\clearpage 

\begin{figure}
\caption{Optical cutouts for the SExtractor Only Catalog.  ID number
(XID in parenthesis) and band is printed in the upper left corner of
each image.  The energy flux in the soft band is in the lower left,
while the flux in the hard 2--10 keV band is in the lower right.  The
iso--intensity contours are at 2, 5, 10 and 20 sigma above the
background.  The box indicates an optical counterpart candidate.  The
positional error is indicated by the circle in the
center.\label{scuts}}
\end{figure}

\clearpage

\begin{figure}
\end{figure}

\clearpage

\begin{figure}
\caption{Optical cutouts for the WAVDETECT Only Catalog.  ID number
(XID in parenthesis) and band is printed in the upper left corner of
each image.  The energy flux in the soft band is in the lower left,
while the flux in the hard 2--10 keV band is in the lower right.  The
iso-intensity contours are at 2, 5, 10 and 20 sigma above the
background.  The box indicates an optical counterpart candidate.  The
positional error is indicated by the circle in the
center.\label{wcuts}}
\end{figure}

\clearpage

\end{document}